\documentclass[aps,prl,twocolumn,preprintnumbers,amsmath,showpacs,floatfix,nofootinbib]{revtex4-1}
\usepackage{graphicx}

\newcommand{\dd}{{\rm d}}

\begin{document}

\preprint{BI-TP 2012/018}

\title{Sequential suppression of quarkonia and high-energy nucleus--nucleus
  collisions}

\author{Nirupam Dutta} \email{nirupam@physik.uni-bielefeld.de}
\author{Nicolas Borghini} \email{borghini@physik.uni-bielefeld.de}
\affiliation{Fakult\"at f\"ur Physik, Universit\"at Bielefeld, 
  Postfach 100131, D-33501 Bielefeld, Germany}

\date{\today}

\begin{abstract}
  According to the usual application of the sequential-suppression picture to 
  the dynamics of heavy quarkonia in the hot medium formed in ultrarelativistic
  nuclear collisions, quark--antiquark pairs created in a given bound or unbound
  state remain in that same state as the medium evolves. 
  We argue that this scenario implicitly assumes an adiabatic evolution of the 
  quarkonia, and we show that the validity of the adiabaticity assumption is 
  questionable. 
\end{abstract}

\pacs{25.75.Nq, 12.38.Mh, 14.40.Pq}

\maketitle

More than 25 years ago, finite-temperature gauge-theory studies on the lattice 
of the screening of static (color) charges prompted Matsui and Satz to suggest 
that in a quark--gluon plasma (QGP) the formation of bound charmonia may be 
prevented~\cite{Matsui:1986dk}. 
It was quickly realized that the various $c\bar c$ or $b\bar b$ states might in 
fact be dissociated at different temperatures~\cite{Karsch:1987pv,Digal:2001ue}.
Over the years, paralleling progress in theoretical studies of quarkonia 
properties (see Refs.~\cite{Rapp:2009my,Brambilla:2010cs} for recent reviews), 
this led to the ``sequential-suppression'' picture of heavy quarkonia as QGP 
thermometer~\cite{Satz:2006kba}.
According to the latter, a given state will be totally suppressed above a 
threshold temperature---which might actually be smaller than the transition
temperature to a QGP. 

This prediction is supported by several approaches. 
First, spectral functions are extracted from lattice-QCD computations of 
correlators of quantum numbers for heavy quark--antiquark pairs~\cite{%
  Umeda:2002vr,Asakawa:2003re,Datta:2003ww,Aarts:2011sm}. 
The disappearance of a peak in the spectral function then signals the 
suppression of a state. 
Alternatively, one resorts to an effective in-medium quark--antiquark 
($Q\bar Q$) potential---either derived from lattice-QCD computations of spatial 
correlators~\cite{Matsui:1986dk,Karsch:1987pv,Digal:2001ue,Wong:2004zr,
  Cabrera:2006wh,Alberico:2006vw,Mocsy:2007yj,Rothkopf:2011db} or derived 
within finite-temperature field theory~\cite{Laine:2006ns,Beraudo:2007ky,%
  Brambilla:2008cx}---, which then enters a Schr\"odinger or Bethe--Salpeter 
equation, whose bound states model the quarkonia in the medium. 
The suppression of a given state takes place when it is no longer bound by the 
potential, although the precise criterion for dissociation might be open to 
discussion~\cite{Mocsy:2008eg}.

In either description, the in-medium bound states of heavy quark--antiquark
pairs are, be it explicitly stated or not, {\it eigenstates\/} of a 
Hamiltonian. 
Note that as the effective potential might actually possess an imaginary 
part~\cite{Laine:2006ns,Rothkopf:2011db,Beraudo:2007ky}, these eigenstates are 
not necessarily stable, but might have only a finite lifetime.
That notwithstanding, the generally accepted picture is that of a 
temperature-dependent suppression pattern, in which at a given energy density of
the medium some states survive, while more excited ones are not bound and thus 
do not form. 

This picture is seemingly supported by experimental measurements in 
ultrarelativistic nucleus--nucleus collisions. 
At the SPS, the NA50 collaboration reported that the anomalous suppression of the 
$\psi'$ in Pb--Pb collisions sets in at a smaller average in-medium path crossed
by the $c\bar c$ pair than for the $J/\psi$~\cite{Alessandro:2006ju}. 
At the much higher LHC energy, the CMS collaboration studied bottomonia and
observed yields consistent with the idea that the excited $\Upsilon(2S)$ and 
$\Upsilon(3S)$ states are more suppressed than the deeper bound 
$\Upsilon(1S)$~\cite{Chatrchyan:2011pe}. 

The usual explanation for such observations in high-energy nucleus--nucleus
collisions is the following, where for the sake of simplicity we leave aside 
so-called ``initial-state effects''.%
\footnote{When comparing {\it relative\/} yields of different states of a given 
  system, say $S$-channel charmonia or bottomonia, for a fixed type of nuclear 
  collisions, these effects should play a minor role.}
At an early stage after the collision, say some instant $t_0$, the created deconfined medium reaches 
high enough an energy density that a given quarkonium state, which we shall refer 
to as ``excited'', is suppressed, while another state of the same system, 
hereafter the ``ground state'', is bound. 
The common lore is then that, as the medium expands and cools down ($t>t_0$), the ground 
state stays unaffected, whereas the depopulated excited state remains 
suppressed, even when the medium temperature has dropped below its 
dissociation threshold. 
The only possibility left to the excited state for being recreated is at the 
transition to the hadronic phase, through the ``recombination'' of till then 
uncorrelated heavy quarks and antiquarks~\cite{BraunMunzinger:2000px,%
  Thews:2000rj}. 
That scenario constitutes the standard implementation of the idea of sequential 
suppression of heavy quarkonia in heavy-ion collisions. 
Note that this description totally ignores the possible finite lifetime of the ground state due 
to the imaginary part of the in-medium potential. 

Inspecting the scenario sketched above critically, it relies on two basic ingredients.
There is first the sequential-suppression pattern in the ``initial condition'' at $t_0$,
whose theoretical foundations we discussed above. 
The second element in the scenario, which to our knowledge has not been examined
before, is the implicit assumption that ``the quarkonium ground state remains 
the ground state'' over the duration of the medium evolution.
Recasting this statement more mathematically, a $Q\bar Q$ pair initially in the 
eigenstate with lowest energy of the (effective) Hamiltonian describing
in-medium quarkonia remains in the lowest-energy eigenstate. 
More generally, the same will hold for every initially bound state---up to late 
electroweak decays which take place outside the medium. 
That is, it is assumed that heavy quarkonia are continuously evolving 
eigenstates of an adiabatically changing instantaneous Hamiltonian. 
Accordingly, the scenario for the sequential suppression of quarkonia in the 
medium created in high-energy nucleus--nucleus collisions relies on the 
hypothesis that the effective in-medium quark--antiquark potential varies slowly
enough that each $Q\bar Q$ pair is at every successive instant in an energy 
eigenstate. 
We now wish to investigate the validity of this assumption. 

Before going any further and to dispel any confusion, let us note that the 
adiabaticity we discuss in this note is neither that of the medium evolution%
---related to the production of entropy---, nor the adiabatic assumption \`a la 
Born--Oppenheimer which allows one to separate gluons from the nonrelativistic 
heavy quarks when writing down an effective potential for the latter~\cite{%
  Bali:2000gf}.

Let $|n(t)\rangle$ denote the eigenstates of an instantaneous Hamiltonian 
${\sf H}(t)$, with respective energies $E_n(t)$. 
Following the approach of Ref.~\cite{Aharonov-Anandan:1987}, a common criterion
for the validity of the adiabatic theorem is the requirement that for every pair
of states $|n(t)\rangle$, $|n'(t)\rangle$ and at every instant in the evolution%
\footnote{We use a system of units in which $\hbar = c = 1$.}
\begin{equation}
\frac{\big|\langle n'(t) | \,\dot{\sf H}(t) \,| n(t) \rangle\big|}%
  {\big[E_n(t)-E_{n'}(t)\big]^2} \ll 1, 
\label{adiabatic-criterion}
\end{equation}
with $\dot{\sf H}(t)$ the time derivative of the Hamiltonian. 
In the case of interest in this note, $\dot{\sf H}(t)$ coincides with the time 
derivative $\dd{\sf V}/\dd t$ of the effective $Q\bar Q$ potential. 
In turn, the latter is simply the product of the rate of change $\dot{T}$ of 
the medium temperature times the derivative $\dd{\sf V}/\dd T$ of the potential 
with respect to $T$, where for the sake of simplicity we have assumed that the 
medium is (locally) thermalized. 
For $\dot{T}$, we took the results from a simulation of central Pb--Pb 
collisions at the LHC within dissipative hydrodynamics~\cite{Shen:2011eg}, 
considering the evolution of temperature at the center of the fireball: 
within the first 7~fm/$c$ of the evolution (that is, as long as $T>200$~MeV), 
$\dot{T}$ always remains larger than about 30~MeV per fm/$c$ and up to 50~MeV 
per fm/$c$ in the early stages. 

For the $Q\bar Q$ potential, we considered the lattice QCD results of 
Ref.~\cite{Kaczmarek:2005ui}, including the parameterization
\begin{equation}
V(r) \sim \frac{\frac{4}{3}\alpha_s(T)}{r}\,
  {\rm e}^{-A\sqrt{1+N_f/6}\,Tg_{2\,\rm loop}(T)\,r},
\label{V_QQ_Olaf}
\end{equation}
with $A\simeq 1.4$ and $g_{2\,\rm loop}(T)$ the 2-loop perturbative coupling, 
where $0.5\lesssim \alpha_s(T)\lesssim 1.0$. 
One then finds as typical amplitude for a matrix element of $\dd{\sf V}/\dd T$ 
between eigenstates of the instantaneous Hamiltonian 
\[
\bigg|\left\langle n'(t) \Big| \,\frac{\dd{\sf V}}{\dd T} \,\Big| n(t) 
  \right\rangle\bigg|\approx 200-500~\text{MeV}\!\cdot\text{fm}.
\]
The numerator in Eq.~\eqref{adiabatic-criterion} is thus of order
(80--160~MeV)$^2$.
In turn, the denominator is of order (100--350~MeV)$^2$ for the excited 
$b\bar b$ states, so that the ratio can be in some cases smaller than 0.1, for
other channels larger than 1. 
Because of those channels, it is far from warranted that the adiabaticity 
assumption holds: 
the potential evolves so quickly that a quark--antiquark pair which at some time
is in a given instantaneous eigenstate will a short while later no longer be in 
the evolved eigenstate, but will have components over all the new 
eigenstates---including the new ground state, which shows that even if 
criterion~\eqref{adiabatic-criterion} holds for the latter, yet it is populated 
by contributions from excited states. 

We wish to emphasize here that this ``repopulation'' mechanism is neither the 
customary recombination at hadronization, nor the feed-down from late decays, 
but a natural consequence of the ``reshuffling'' of $Q\bar Q$ states due to the 
rapid medium evolution.

A naive picture of the effect of this rapid evolution is provided by dividing 
the typical size $r_{\rm rms}\approx 0.3$--0.75~fm of a bound bottomonium by the
characteristic velocity $v\sim 0.3c$ of the nonrelativistic constituent quark 
and antiquark, which gives a duration $\tau\approx 1$--2.5~fm/$c$ for an 
``orbit'' of the $b$ quark. 
On such a time scale, the QGP cools down by 30 to 75~MeV, resulting in a 
significant change in the effective potential~\eqref{V_QQ_Olaf}, which 
illustrates why the adiabatic evolution of bottomonia is far from being 
warranted. 

As a final argument against using the hypothesis of an adiabatic evolution of 
$Q\bar Q$ pairs in a QGP, we note that recent studies emphasized the fact that 
even when criterion~\eqref{adiabatic-criterion} is satisfied---i.e., the 
evolution is slow---, the system with evolving Hamiltonian can be driven from 
one instantaneous eigenstate to a different one at later times by resonant 
interactions (see e.g.\ Ref.~\cite{Amin:PRL2009}). 
The latter lead to Rabi oscillations between eigenstates---that is, they are 
tailored to induce transitions which violate the adiabatic theorem---on a time 
scale given by the inverse of the Rabi frequency $\omega_R$.

In the case of a $Q\bar Q$ pair in a quark--gluon plasma at the temperatures 
found in high-energy nuclear collisions, there are obviously plenty of degrees 
of freedom around with energies matching possible transition lines. 
The corresponding Rabi frequencies however depend on the interaction term. 
Adopting for the sake of illustration a dipolar interaction, one finds values 
of $\pi/\omega_R$, which in a two-level system is the time after which a 
transition has occurred with probability 1, of the order 2 to 20~fm/$c$, 
depending on the transition Bohr frequency, the medium size and the assumed 
coupling strength. 
This means that on such a time scale a $Q\bar Q$ pair certainly does not remain 
in the same instantaneous eigenstate, which again hints at the invalidity of 
the adiabatic theorem for heavy quarkonia in a dynamical QGP. 

One might be tempted to argue that in an effective-potential approach, the  
transition-inducing degrees of freedom have been integrated out. 
Yet the construction of an effective theory ultimately relies on the adiabatic 
theorem~\cite{Coleman:book}, so that it is inconsistent to use the notion 
blindly here. 
More precisely, we surmise, although we have not investigated this idea in 
detail, that the violation of adiabaticity caused by resonant interactions 
translates into the imaginary part of the effective in-medium potential, which 
physically has the same effect of giving a finite lifetime to the Hamiltonian 
eigenstates. 

In summary, we have shown that the usual scenario for the sequential 
suppression of heavy quarkonia in the hot medium created in ultrarelativistic 
heavy-ion collisions relies on the hypothesis that $Q\bar Q$ pairs evolve 
adiabatically in the medium. 
We have then presented several arguments which make us doubt that this 
assumption holds. 
In our view, this hints at the idea that a given quark--antiquark pair is in a 
constantly changing linear combination of instantaneous energy eigenstates, 
rather than in a smoothly evolving unique eigenstate. 
As a consequence, one should explicitly follow the evolution of the pair in 
the QGP, using a dynamical microscopic description, as attempted in 
Refs.~\cite{Borghini:2011ms,Dutta:inprep}. 
Such an approach should of course be suited for rapidly evolving media, so as 
to eventually be able to compare with experimental results from high-energy
nucleus--nucleus collisions.
On the other hand, in the regime of a (quasi-)static environment of the 
quarkonia it should also make contact with equilibrium-based formalisms as 
lattice QCD or finite-temperature field theory. 

\section{Acknowledgments}

We thank Chun Shen and Ulrich Heinz for providing us with the detailed results 
from their hydrodynamic simulations. 
N.\ D.\ acknowledges support from the Deutsche Forschungs\-gemeinschaft under
grant GRK 881.

\end{document}